\documentclass[a4paper]{article}
\usepackage{wrapfig}
\usepackage{float}
\usepackage[pages=all, color=black, position={current page.south}, placement=bottom, scale=1, opacity=1, vshift=5mm]{background}
\SetBgContents{
	\tt 
}      
\usepackage[utf8]{inputenc}
\usepackage{cite}
\usepackage[margin=1in]{geometry} 

\usepackage{amsmath}
\usepackage{amsthm}
\usepackage{amssymb}

\usepackage[utf8]{inputenc}
\usepackage{hyperref}
\hypersetup{
	unicode,
	pdfauthor={Author One, Author Two, Author Three},
	pdftitle={A simple article template},
	pdfsubject={A simple article template},
	pdfkeywords={article, template, simple},
	pdfproducer={LaTeX},
	pdfcreator={pdflatex}
}

\usepackage[sort&compress,numbers,square]{natbib}

\theoremstyle{definition}

\usepackage{graphicx, color}
\graphicspath{{fig/}}

\usepackage{algorithm, algpseudocode} 
\usepackage{mathrsfs} 

\usepackage{lipsum}

\title{Optimizing Sequence Alignment with Scored NFAs}
\author{Ryan Karbowniczak \and Rasha Karakchi}

\date{
	University of South Carolina\\ 
    \texttt{karbownr@email.sc.edu}\\
    \texttt{karakchi@cec.sc.edu}\\%
	[2ex]%
}

\begin{document}
	\maketitle
	
	\begin{abstract}
The rapid increase in symbolic data has underscored the significance of pattern matching and regular expression processing. While nondeterministic finite automata (NFA) are commonly used for these tasks, they are limited to detecting matches without determining the optimal one. This research expands on the NAPOLY pattern-matching accelerator by introducing NAPOLY+, which adds registers to each processing element to store variables like scores, weights, or edge costs. This enhancement allows NAPOLY+ to identify the highest score corresponding to the best match in sequence alignment tasks
through the new-added arithmetic unit in each processor element. The design was evaluated against the original NAPOLY, with results showing that NAPOLY+ offers superior functionality and improved performance in identifying the best match.

The design was implemented and tested on zynq102 and zynq104 FPGA devices, with performance metrics compared across array sizes from 1K to 64K processing elements. The results showed that memory usage increased proportionally with array size with Fmax decreasing as the array size grew on both platforms. The reported findings focus specifically on the core array, excluding the impact of buffers and DRAMs.
 \end{abstract}
	
	\section{Introduction}

With the rapid expansion of symbolic data in fields such as internet data, biological data, and financial data, the need for efficient pattern matching and regular expression processing has increased \cite{Bradshaw2016, Karakchi2016, Karakchi2023, Yu2017}. Some pattern matching computations are modeled as nondeterministic finite automata (NFA) \cite{Karakchi2017a,Karakchi2019}. 
NFAs can have multiple states activated simultaneously, allowing concurrent operation of multiple next-state functions. This requires memory bandwidth that scales with state activation rates, which often results in memory bottlenecks when they are executed on general-purpose platforms. 

\begin{figure}[h]
    \centering
\includegraphics[width=0.75\linewidth]{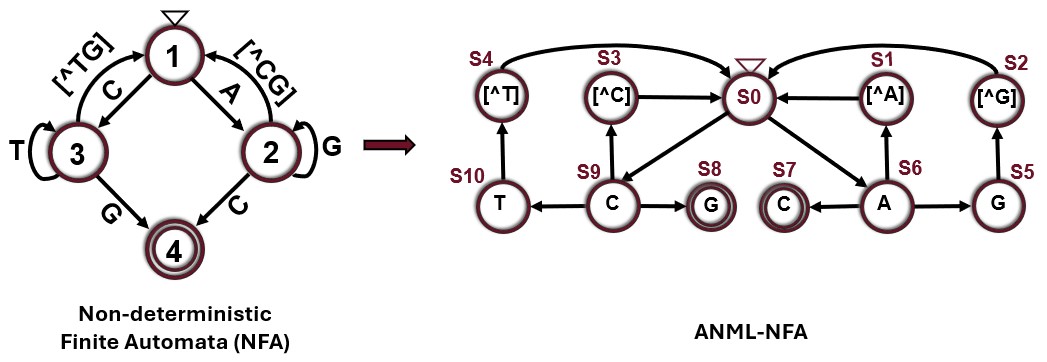}
    \caption{Example of NFA-ANML conversion}
    \label{fig:nfa-anml}
\end{figure}

This inefficiency has driven interest in Domain-Specific Architectures (DSA) that can effectively utilize NFA parallelism. Many current automata processors are built using specialized hardware such as FPGAs \cite{Karakchi2023} and ASICs \cite{Roy2014, Wang2017}, which enable efficient parallel processing. These specialized processors are faster than general-purpose CPUs for specific tasks and offer better energy efficiency and cost-effectiveness compared to traditional computing methods.


Recent studies show that while current automated processors excel at detecting pattern matches, many modern applications demand the ability to determine the optimal match path among various alternatives. This requirement is evident in areas such as graph-based shortest-path algorithms, probabilistic modeling, and DNA sequence alignment \cite{Bradshaw2016,Yu2017}. For example, in biology, the similarity between aligned sequences is often measured using an additive scoring function with penalties for gaps. In the example illustrated in the left part of Figure \ref{fig:nfa-anml}, a DNA data set is represented as a non-deterministic finite automaton (NFA). Using match, mismatch, and gap scores of +2, -1, and -2, respectively, the highest alignment score is 6 for the sequence "AGC" and -1 for "AGATG."

The primary goal of this work is to enhance the existing FPGA-based NAPOLY (nondeterministic finite automata overlay) to identify and report the best sequence alignment by determining the highest score. Our method is based on finite weighted automata \cite{Wikipedia2024, Mohri2009, Khoussainov2020}, integrating weights (scores) into automata transitions to compute the accumulated highest score and evaluate the quality of the output \cite{Karbowniczak2024, Mohri2009, Mohri2012}.

Section 2 provides an overview of NAPOLY and the representation of nondeterministic automata, along with background information on weighted finite automata and the Viterbi algorithm for identifying the best match. Section 3 details the proposed approach and its architectural design. Section 4 presents the experimental results and performance, and Section 5 concludes the work with a summary and discussion of future plans.

\section{Background}

\subsection{NAPOLY Automta Processor}
NAPOLY is a flexible and reusable architecture specifically designed to implement nondeterministic finite automata (NFA) descriptions. It introduces constraints such as hardware fan-out, which determines the maximum number of outgoing transitions per State Transition Element (STE), as well as the maximum allowable distance between connected elements \cite{Karakchi2023, Bakos2023, Karakchi2017a}.

Each STE incorporates an OR-gate that aggregates activation signals from up to f predecessor STEs, where f is defined by the hardware fan-out. An STE activates its state bit in the next clock cycle if it detects at least one incoming activation signal while simultaneously receiving a one-bit signal from the state table. If these conditions are not met, the state bit resets unless the "start" bit is enabled.

\begin{figure}[h]
    \centering
\includegraphics[width=0.8\linewidth]{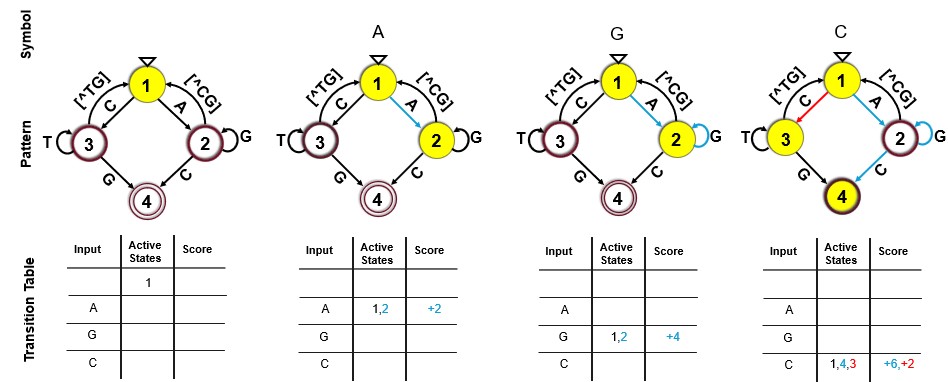}
    \caption{Tracking multiple paths simultaneously}
    \label{fig:challenges}
\end{figure}
When active, an STE transmits activation signals to its f outputs. These signals are combined with "interconnect configuration" bits via AND gates before being routed to the OR gates of the successor STEs. The configuration bits define the edges (connections) between states mapped onto STEs, forming a point-to-point programmable interconnect. Each wire and its associated configuration bit establish a specific link between the STEs.
The STEs on the FPGA are physically connected using point-to-point links, enabling each STE to communicate with itself and up to f-1 neighboring STEs. A one-dimensional addressing scheme is used to identify each STE with an ID n, allowing it to connect to successor STEs within the range n - [(f - 1)/2] to n + [f / 2].

NAPOLY utilizes the Automata Network Markup Language (ANML) format, an alternative representation of NFAs developed by Micron \cite{Dlugosch2014}. In ANML, transition labels are associated with states rather than edges. For example, Figure \ref{fig:nfa-anml} shows how a DNA NFA is converted into its ANML representation, where the transitions in the original NFA correspond to states or nodes in the ANML version.

\textbf{Limitations of NAPOLY}: As illustrated in Figure \ref{fig:challenges}, NAPOLY assumes that the start state (state 1) remains continuously active. This design results in multiple states being active simultaneously, creating redundant paths to the same destination state (e.g., state 4).

For example, when the input symbol is "C", states 1, 3, and 4 become active concurrently: state 1 remains active as the start state, state 3 is triggered by matching "C" along the red path, and state 4 is reached by matching "C" through the blue path.

Although NAPOLY reports the first match it detects, it does not account for the path cost. This limitation poses a significant challenge for applications such as sequence alignment, where identifying the highest scoring path is essential for ensuring accurate and high-quality alignments.

\subsection{Weight Finite Automata}

Weighted finite automata (WFAs) extend traditional finite automata (FAs) by assigning weights to transitions, states, or both. Unlike standard finite automata that only recognize strings by moving between states, WFAs incorporate additional quantitative information on their transitions. These weights can represent attributes such as costs, probabilities, or scores, allowing the automata to generate quantitative outputs \cite{Mohri2009, Mohri2012, Khoussainov2020,Suresh2021,  Ng2015}. WFAs are capable of calculating metrics like the probability of accepting a string (in probabilistic automata) or the minimal cost for acceptance (in cost-based automata).

Deterministic Weighted Automata (DWA) have at most one transition for each input symbol per state, offering better computational efficiency. In contrast, nondeterministic Weighted Automata (NWA) allow multiple transitions for the same input symbol, providing greater flexibility but at the cost of efficiency.

A \emph{Weighted Finite Automaton} (WFA) is a tuple \( (Q, \Sigma, T, q_0, F, w) \) where: 
\begin{itemize}
    \item \( Q \) is a finite set of states.
    \item \( \Sigma \) is a finite alphabet.
    \item \( T \subseteq Q \times \Sigma \times Q \) is the transition relation.
    \item \( q_0 \in Q \) is the initial state.
    \item \( F \subseteq Q \) is the set of final states.
    \item \( w: T \to \mathbb{R} \) is the weight function that assigns a weight to each transition.
\end{itemize}

\section{Proposed Design}

\begin{figure} [h]
    \centering
    \includegraphics[width=0.5\linewidth]{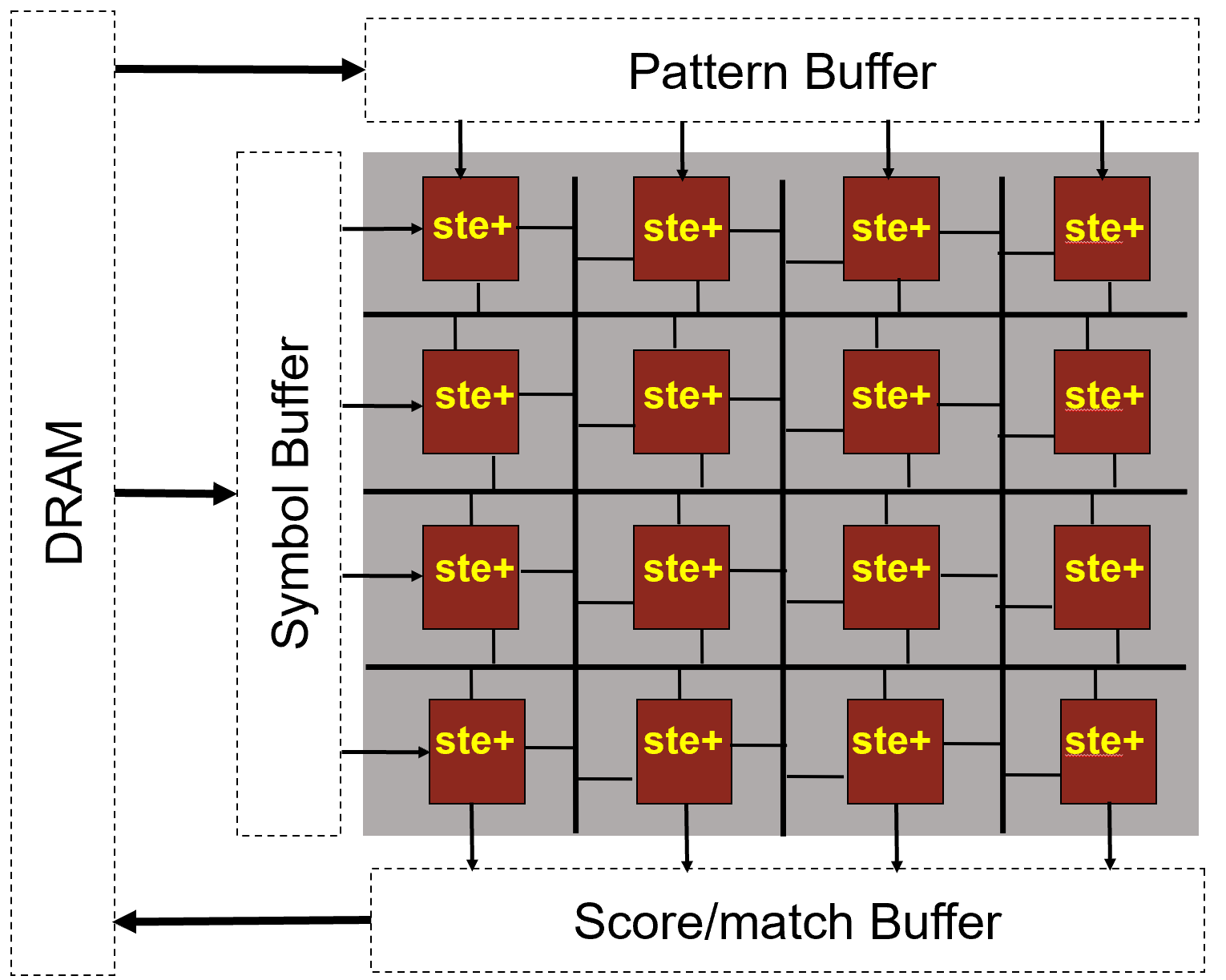}
    \caption{NAPOLY+ layout of STE+ within the design}
    \label{fig:layout}
\end{figure}

The NAPOLY+ design consists of a two-dimensional array of specialized STEs (STE+) which incorporate symbol, score (weight), and arithmetic components to accumulate scores along the path and determine the final score \cite{karbowniczak2024scored}. Figure \ref{fig:layout} depicts the array which intuitively resembles the layout of the Configurable Logic Blocks that make up an FPGA.

The design of each STE+ preserves the fundamental logic of NAPOLY's STE design \cite{Karakchi2016, Karakchi2019, Karakchi2020, Karakchi2023, Karbowniczak2024}. Each STE+ still combines activation signals from all f predecessors and will reset its state bit unless the "start" bit has been set. Furthermore, in the case of the state bit being set, (STE+)s similarly broadcast an activation signal to each of their f outputs that is AND’ed against a corresponding “interconnect configuration" bit before potential successors are checked and the process is repeated.

\begin{figure}[h]
    \centering
    \includegraphics[width=0.75\linewidth]{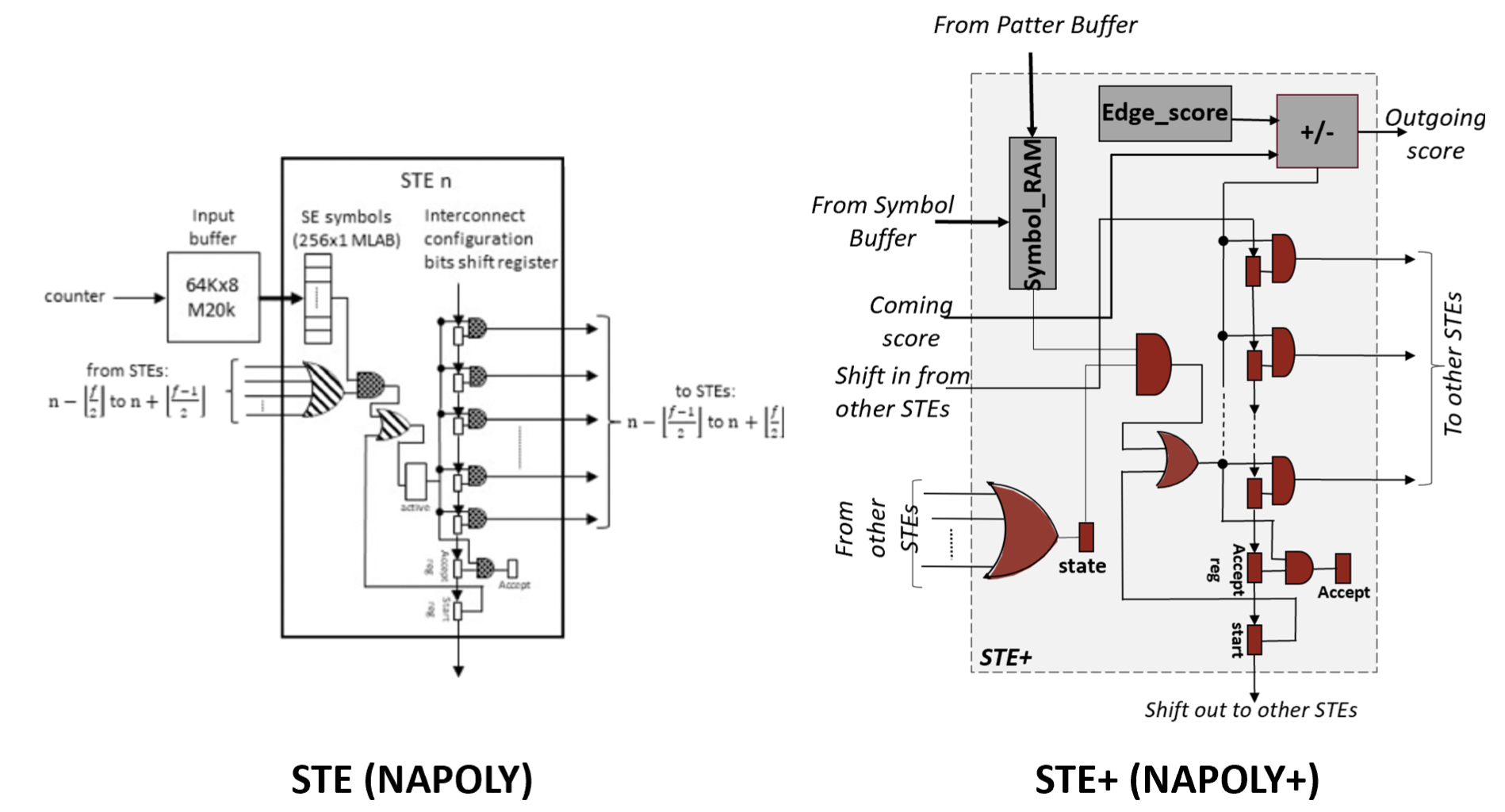}
    \caption{The original STE design compared to the STE+ design of NAPOLY+}
    \label{fig:design}
\end{figure}

While maintaining the original logic, the STE+ simultaneously calculates an outgoing score, as depicted on the right of Figure \ref{fig:design}. This score is based on the incoming score from its predecessor and its edge score, a fixed number stored in a register and configured during the reconfiguration stage. To handle the possibilities of mismatches and gaps, we designed all (STE+)s to be connected to the start STE+. However, this implementation is constrained by the array size and the maximum fan-out which is limited by the interconnection design and hardware resources. The interconnections are structured as a grid of global wires and local wires on the fine-grained layer of the chip. The horizontal wires are limited by the maximum bus size that can be created on the selected FPGA (1 million wires), while the vertical wires represent the local fan-out wires connected to and from the STE+.

During the operation, start STE+ remains active at all times, enabling the processing of multiple symbols simultaneously and allowing multiple STE+ to be active at once. To distinguish between new symbol signals and mismatching incoming signals, we initialize the incoming scores to zero for every new symbol, indicating the possibility of a new path. We assigned a specific fan-in signal exclusively for the new symbol. The accepting (STE+)s are designed similarly to regular STE but have no connection with the start state since a match is found and the score is reported once the accepting STE+ is activated. Both pattern sets and input sequences are stored in buffers, while the vector of accepting state IDs, input symbol offsets, and scores are flushed out to an output buffer. All buffers are assumed to be connected to DRAMs.

\section{Experimental Results}

This research focused on evaluating the hardware efficiency and performance of the NAPOLY+ design on two FPGA devices, comparing its results to those of its predecessor, the NAPOLY design. We implemented the design on two Zynq UltraScale+ MPSoC FPGA devices, optimized for high-performance applications. 

The first device, ZCU104, features approximately 504K Logical Cells (LCs), 461K Flip-Flops (FFs), and 6.2 Mb Mb of distributed memory \cite{Xilinx2019a}. The second device, ZCU102, includes roughly 600K LCs, 550K FFs, and 8.8 Mb of distributed memory \cite{Xilinx2019b}. 
As detailed earlier, NAPOLY+ is an enhanced version of NAPOLY that incorporates an arithmetic unit and a register to store the local score for each processor element.

\begin{figure}
    \centering
    \includegraphics[width=0.5\linewidth]{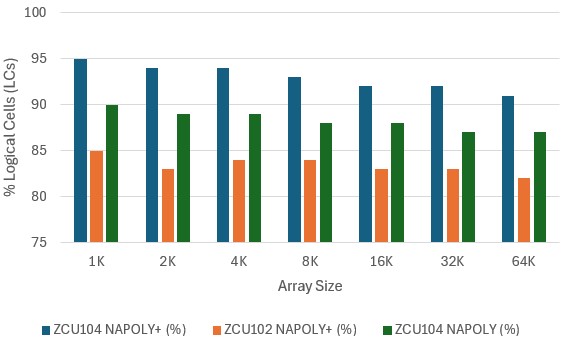}
    \caption{Percentage of Logical Cells utilized across each design configuration}
    \label{fig:luts}
\end{figure}

\textbf{Hardware resources versus array size:} Figure \ref{fig:luts} illustrates a notable increase in Logical Cells (LCs) utilization between the NAPOLY+ and NAPOLY implementations, with NAPOLY+ achieving utilization rates of 90–95\%, compared to 87–90\% for the original NAPOLY design. On the ZCU102 board, NAPOLY+ maintained a utilization rate just under 85\%. This difference is attributed to the added logic in NAPOLY+ for determining the best matches. The higher LC utilization was anticipated and remains within the hardware's capacity.

\begin{figure} [h]
    \centering
    \includegraphics[width=0.5\linewidth]{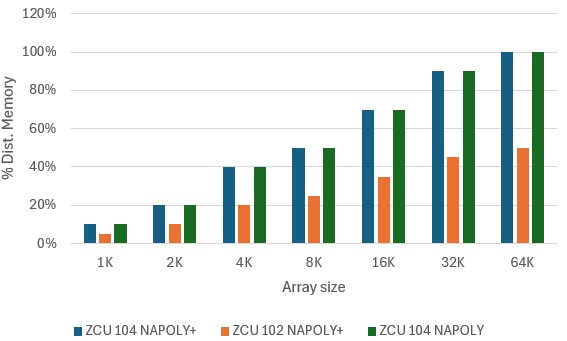}
    \caption{The distributed memory usage for each design configuration}
    \label{fig:distributed}
\end{figure}

As explained in the previous section, the distributed memory (fast scratchpad memory \cite{Karakchi2024}) is used to store the label or symbol of each processor element, which is consistent in both the extended version (NAPOLY+) and the original version (NAPOLY). Consequently, as shown in Figure \ref{fig:distributed}, the results for NAPOLY and NAPOLY+ are similar when tested on the ZCU104 board. However, since the ZCU102 features a larger amount of distributed memory, the percentage of memory utilized by NAPOLY+ on the ZCU102 is lower. This provides the ZCU102 with a greater capacity to support larger array sizes exceeding 64K.

\begin{figure}[h]
    \centering
    \includegraphics[width=0.5\linewidth]{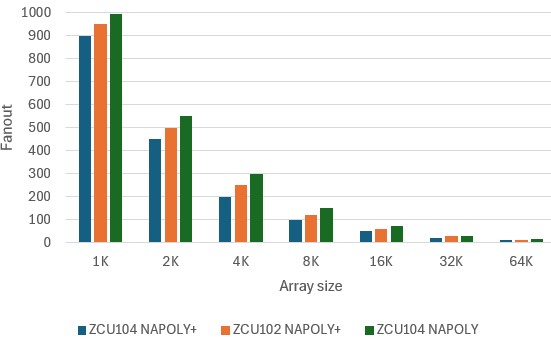}
    \caption{Maximum fanout allowed for each design configuration}
    \label{fig:fanout}
\end{figure}

\textbf{Array size versus Fanout:} Figure \ref{fig:fanout} shows that the ZCU104 consistently has lower fan-out than the ZCU102 for NAPOLY+ across all array sizes. This can be because ZCU102 is slightly larger than ZCU104 in terms of the hardware resources. 

NAPOLY+ also reduces fan-out in comparison to the original NAPOLY. For example, at 1K, NAPOLY + in ZCU104 reduces fanout by about 10\% compared to ZCU102 and the original NAPOLY in ZCU104. As array sizes grow, the fan-out values for both boards and implementations become similar, with little difference at 16K and above. This shows that NAPOLY+ is most effective at smaller scales but still performs well as the size increases.

\textbf{Performance: }Figure 8 presents a detailed comparison of two FPGA boards, the ZCU104 and ZCU102, evaluating
their maximum frequency (Fmax) and performance (throughput in MB/sec) across various array sizes.
\begin{figure} [h]
    \centering
\includegraphics[width=0.5\linewidth]{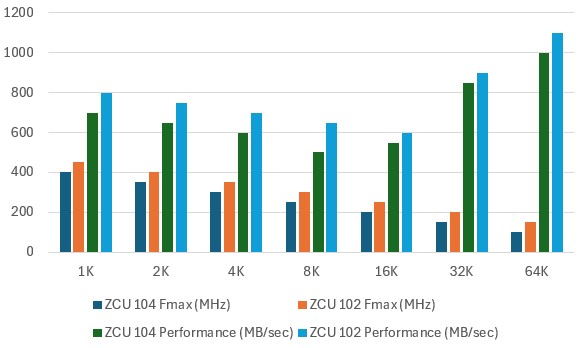}
    \caption{Maximum frequency and Performance of NAPOLY+ on ZCU104 and ZCU102 FPGA boards}
    \label{fig:performance}
\end{figure}
The data show a consistent trend where both boards experience a decrease in Fmax as the array
size increases, which is expected due to the increased complexity of computation and data handling.
However, ZCU102 outperforms ZCU104 in both Fmax and throughput across all array sizes, with the
performance gap widening slightly as the array size grows.
The difference in Fmax between the two boards is noticeable, with ZCU102 being faster by
50 MHz at the 1K array size and up to 100 MHz faster at the 64K array size. This suggests that ZCU102 is better optimized for handling higher clock speeds, which could contribute to its superior overall
performance.

\section{Conclusion and Future Work}

In this study, we introduced NAPOLY+, an enhanced version of the NAPOLY pattern matching accelerator, designed to address the limitations of traditional nondeterministic finite automata (NFA) by incorporating register-based score tracking. This improvement allows for the identification of the best match in sequence alignment applications, particularly in biological sequence analysis. Experimental results on zynq102 and zynq104 UltraScale+ FPGA devices demonstrated that NAPOLY+ outperforms the original NAPOLY in terms of functional capabilities, providing a more efficient solution for pattern matching tasks. 

Specifically, while both Zynq devices exhibited a proportional increase in memory usage with increasing array size, the zynq102 demonstrated a higher maximum frequency (Fmax) and greater overall performance compared to the zynq104. These results highlight the advantages of leveraging more powerful FPGA platforms for larger and more complex pattern matching applications.

For future work, we plan to further optimize the NAPOLY+ architecture by exploring more advanced interconnection strategies and expanding the processing element array sizes. In particular, investigating the impact of various memory hierarchies, including DRAM and buffers, on overall performance will provide valuable insights into improving efficiency.

Additionally, we aim to extend the scope of our design to other applications beyond sequence alignment, such as machine learning and graph processing, to further demonstrate the versatility of the NAPOLY+ accelerator. Future comparisons between different FPGA platforms, including more recent models, will help identify the optimal hardware configurations for large-scale pattern matching and regular expression processing tasks.

\bibliographystyle{plain}
	\bibliography{main}

\end{document}